\documentclass[11pt]{article}
\usepackage{moriond,epsfig}
\usepackage{amsmath}
\usepackage{amssymb}

\newcommand{\as}{\alpha_s}

\newcommand{\tot}{\mathrm{tot}}
\newcommand{\GeV}{\,\mathrm{GeV}}
\newcommand{\TeV}{\,\mathrm{TeV}}
\newcommand{\lan}{\langle}
\newcommand{\ran}{\rangle}
\newcommand{\ext}{\mathrm{ext}}

\def\Journal#1#2#3#4{{#1} {\bf #2}, #3 (#4)}

\def\PLB{{\em Phys. Lett.}  B}

\def\JHEP{{\em JHEP}}

\def\be{\begin{equation}}
\def\ee{\end{equation}}
\def\bea{\begin{eqnarray}}
\def\eea{\end{eqnarray}}

\def\bit{\begin{itemize}}
\def\eit{\end{itemize}}
\begin{document}
\vspace*{4cm}
\title{%
On strategies for determination and characterization of the underlying event}

\author{S. SAPETA}

\address{%
   LPTHE, UPMC Univ. Paris 6 and CNRS UMR 7589\\
   Paris, France
}

\maketitle\abstracts{
We discuss the problem of the separation and description of the underlying
event~(UE) within two existing approaches to UE measurement: the ``traditional''
method, widely used at Tevatron, and a recently proposed jet-area/median
method. A simple toy model of UE is developed in order to understand how
these approaches perform. We find that both methods are comparably good
for measuring average properties of the UE but the jet-area/median
approach is favorable for determining fluctuations. We also use the latter
method to study the UE from several existing Monte Carlo generator tunes.
We investigate which characteristics of the underlying event might be
useful to measure in order to improve understanding of its properties and
to simulate it well. These include transverse momentum density per area,
intra- and inter-event fluctuations and correlations.
}

%%%%%%%%%%%%%%%%%%%%%%%%%%%%%%%%%%%%%%%%%%%%%%%%%%%%%%%%%%%%%%%%%%%%%%%%%%%%%%%%
\section{Introduction}
The hard processes studied at hadron collider are nearly always accompanied by
the \emph{underlying event}~(UE). This predominantly soft activity affects a
wide variety of high-$p_t$ measurements, e.g. by introducing a bias or by
degrading kinematic jet reconstruction.
Therefore, a good understanding and precise theoretical control over the
underlying event is of great importance in order to fully exploit the potential
of LHC.

Study of the UE, both experimental and  theoretical, faces, however, a number of
problems. 
The first of them appears already at the level of definition since the very
concept of the underlying event is ambiguous. 
This is because none of the events measured at hadron colliders has 
a~clear boundary between the hard part and the UE. 
Also modeling of the UE is difficult. Most successful and widely used Monte
Carlo models generate the majority of the underlying activity via the mechanism
of multiple parton interactions. There are however other conceivable mechanisms
which could contribute to the UE and which have not been given as thorough
study. Questions include e.g. the role of correlations as well as the effects of possible contributions from the BFKL type radiation.

Given the above difficulties, one can ask the question if, at least, it would
be possible to measure the UE in, admittedly always to some extent arbitrary,
but well defined and well controlled way. 
%
%This would involve estimation of potential systematic uncertainties introduced
%by a method. This would help constraining, tuning and improving the models

This leads us to addressing the following two questions:~\cite{Cacciari:2009dp}
What do we really measure with existing methods of UE determination
and which observables are interesting to measure?
To answer the first question we develop a simple toy model and use it as a
testing ground for two existing methods of UE measurement. To address the second
question we study the UE from Monte Carlo models and identify a set
of quantities which could serve as valuable characteristics of the UE.

%%%%%%%%%%%%%%%%%%%%%%%%%%%%%%%%%%%%%%%%%%%%%%%%%%%%%%%%%%%%%%%%%%%%%%%%%%%%%%%%
\section{Relevant characteristics and measuring methods of the UE}
We concentrate on quantities related to energy flow. Those
involve the main observable called~$\rho$, which is defined as the amount of
transverse momentum of UE per unit area. We will be also interested in rapidity
dependence of $\rho$, its point-to-point fluctuations within a single event,
$\sigma$, and its fluctuations from event to event as well as the point-to-point
correlations.

Two methods exist which allow one to study the underlying event: the
traditional approach,~\cite{Albrow:2006rt,KarFieldDY} widely used at Tevatron,
and the more recent area/median based
approach.~\cite{Cacciari:2007fd,Cacciari:2008gn} Certain freedom is present in
each of the methods. Below, we describe the version used in our study.

In the traditional method all particles within a certain rapidity range are
first clustered into jets. Then, four regions in the transverse plane are
defined based on the position of the leading jet, whose direction defines
$\phi=0$: the ``towards'' region, $|\phi| < \pi/3$, an away region, $2\pi/3 <
|\phi| < \pi$, and two transverse regions. The characteristic $p_t$ of the UE is
defined as $p_t$ in one or both of those transverse regions.
To further reduce the contamination from the perturbative radiation, the 
transverse regions are labeled, on an event-by-event basis, as TransMin and
TransMax depending on their relative value of $p_t$. The
contribution of perturbative radiation to the average $p_t$ in TransMin should
be suppressed by an additional $\alpha_s$ with respect to the average result
from both regions TransAv.

The area/median method is jet-based, exploits the concept of jet areas
\cite{Cacciari:2008gn} and can be carried out using the {\tt FastJet}
package.~\cite{Cacciari:2005hq,FastJet} It starts by adding a dense set of
infinitely soft particles, \emph{ghosts}, to an event. Then, all particles (real
and ghosts) are clustered with the C/A algorithm~\cite{Cam,Aachen} leading to a
set of jets ranging from hard to soft. The typical UE scale in the event is
defined as
\begin{equation}
  \label{eq:median}
  \rho = \mathop{\mathrm{median}}_{j \in \mathrm{jets}} 
  \left[ \left\{ \frac{p_{tj}}{A_j}\right\}\right]\,,
\end{equation}
where $A_j$ is a jet area, designed to measure susceptibility of a jet to soft
radiation.
In a similar manner,  a quantity measuring the intra-event
fluctuations, denoted as $\sigma$, is determined from the sorted list
$\{p_{tj}/A_j\}$. It is given by the value for which $15.86\% $ of jets have
smaller $p_{tj}/A_j$.  

%With such definition, in the case of Gaussian
%distribution of UE, $34.14\% $ of jets satisfy $\rho - \sigma/\sqrt{A_j} < p_{tj}/A_j < \rho$. 
%

%%%%%%%%%%%%%%%%%%%%%%%%%%%%%%%%%%%%%%%%%%%%%%%%%%%%%%%%%%%%%%%%%%%%%%%%%%%%%%%%
\section{Understanding systematic effects: a toy model study}
It is not guaranteed a priori that the methods described above will give a
sensible result for the characteristic momentum scale of UE. Therefore, to
better understand how those methods perform, we have tested them against a
simple toy model.

The model involves two components: a soft one, which we identify with genuine
UE, and a hard one, which comes from perturbative contamination. The main
parameters of the soft component are: the average density of particles per unit
area, $\nu$, and the average $p_t$ of a particle, $\mu$.  
Therefore, the true value  of $\rho = \nu \mu$ by definition. The number of
particles is governed by the Poisson distribution and the $p_t$ of a single
particle by the exponential distribution. 
The hard contamination is modeled as coming from the initial state
radiation of soft and collinear primary emissions. We assume this radiation to
be independent of rapidity and $\phi$ for $1 \lesssim p_t \lesssim
 Q$, where $Q$ is a hard scale of the process (e.g. half of the hardest jet's
 $p_t$ for the dijet event). 

One of the results of the toy model study are the biases for $\rho$ and $\sigma$
extraction.~\cite{Cacciari:2009dp}
For instance, the hard radiation introduces the bias for $\rho$ determined in
the traditional approach, $\delta \rho \sim \as^2\, Q$, which depends
linearly on the hard scale of the process. 
The corresponding bias in the area/median approach has only $\log(\log)$
dependence 
$\delta\rho \sim \sigma 
\left(n_b/A_\tot + \text{const}\cdot \log \log Q\right)$, 
%\left(n_b/A_\tot + \text{const}\cdot \ln \left[\as(Q_0)/\as(Q)\right]\right)$, 
%
where $n_b$ is the number of final state born particles and $A_\tot$ is
the total used area of the event. 

Another important outcome of the study with the toy model concerns fluctuations.
The extracted values of $\rho$ vary from one event to another, even if the same
$\rho$ distribution is used to generate all events. 
This is because, usually, one works only with limited part of an event ($y$~and $\phi$ cuts) and a finite number of objects (particles, jets).
The magnitude of these intrinsic event-to-event fluctuations is an important
characteristic of a method.

\begin{figure}[t]
\psfig{figure=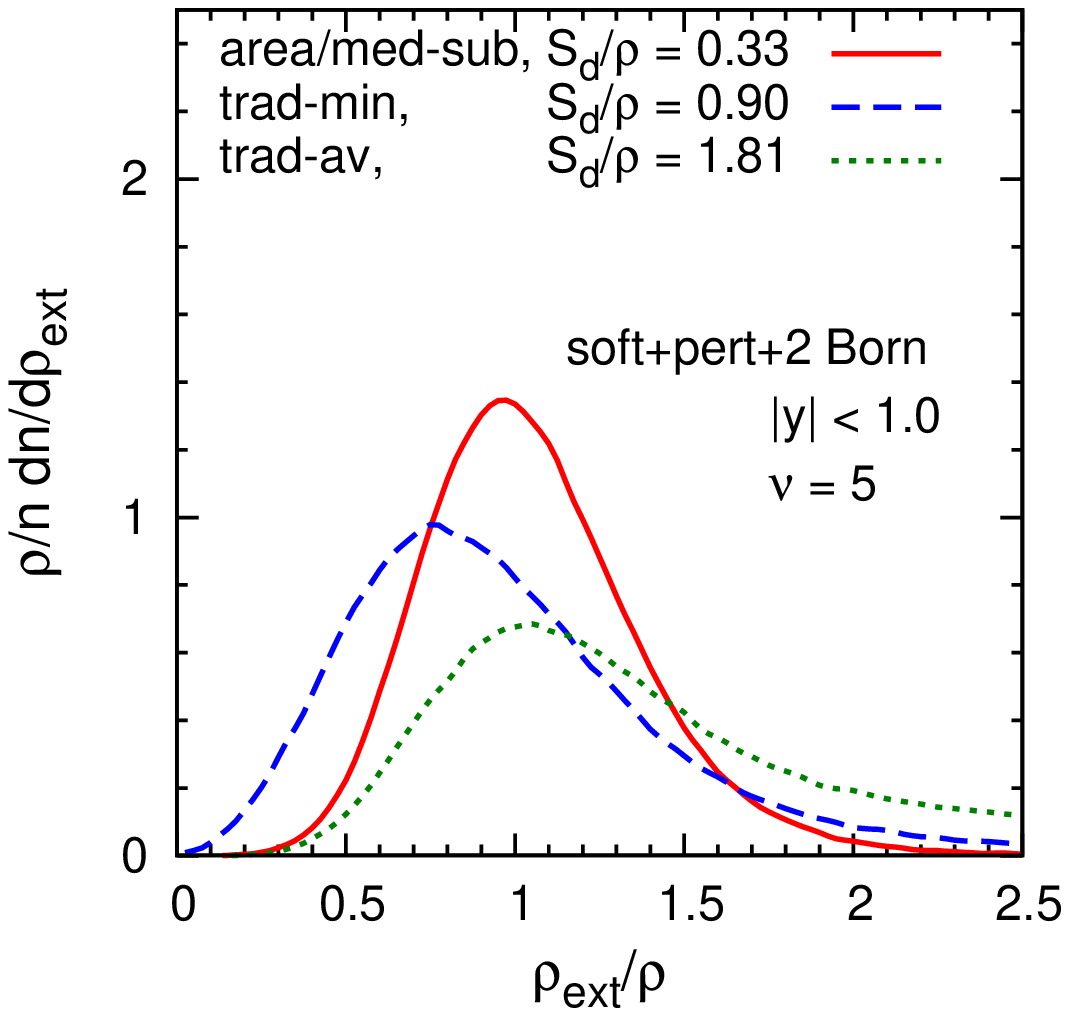,
       height=7.4cm,angle=-90}
\hfill
\psfig{figure=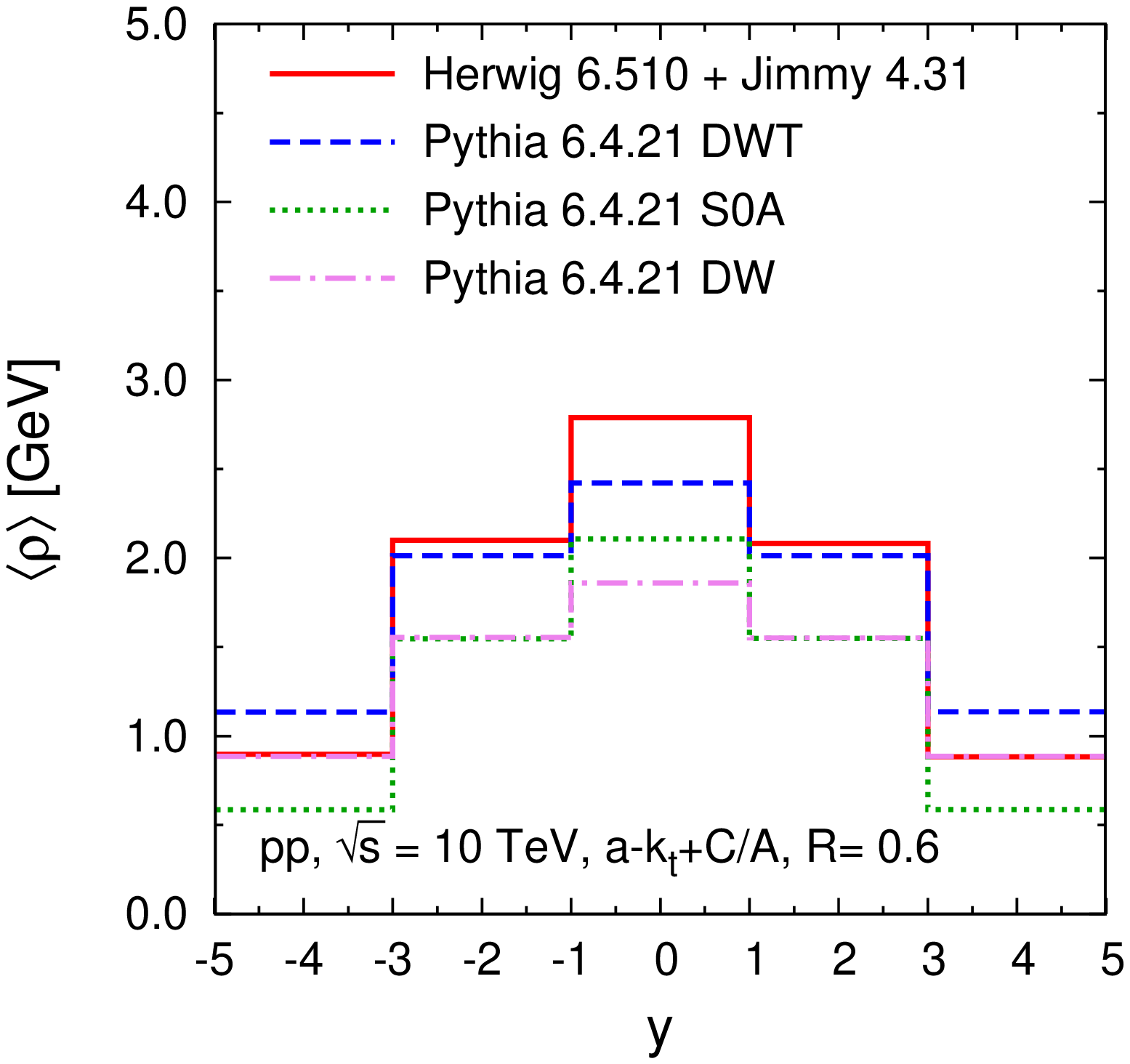,
       height=7.4cm,angle=-90}
\caption{
    Left: Distribution of $\rho_\ext$ from toy model.
    Right: Rapidity dependence of $\lan \rho \ran$ from series of MC model.
\label{fig:rhohist-rhomc}}
\end{figure}

Fig.~\ref{fig:rhohist-rhomc} (left) shows the histograms of $\rho$ from the toy
dijet events extracted with the traditional method in its TransMin and TransAv
variants and with the area/median method. The latter was used after removing
the two hardest jets from the list of jets. 
We see that using the area/median
based method results in the peak which is better both in terms of position and
the width. The real difference is seen, however, in the values of the standard
deviation. In the traditional method, it comes out as big as the value of
$\rho$ itself, while staying moderately small for the area/median method. The
reason for this is again related to the difference in the sensitivity to the
hard radiation between average and median (linear vs $\log(\log)$ as discussed
above for~$\rho$).

%%%%%%%%%%%%%%%%%%%%%%%%%%%%%%%%%%%%%%%%%%%%%%%%%%%%%%%%%%%%%%%%%%%%%%%%%%%%%%%%
\section{Approaching real life: Monte Carlo study}
The weak sensitivity of the area/median method to the hard radiation makes it
advantageous for event-by-event studies and for measuring fluctuations.
Therefore, we have used this method to examine the underlying event from a
series of Monte Carlo generators and tunes.~\cite{Cacciari:2009dp}
We have carried out the study for dijet events at $\sqrt{s} = 10\TeV$. The
leading and next to leading jets, found with the anti-$k_t$
algorithm~\cite{Cacciari:2008gp} with
$R=0.6$, were required to lie in the rapidity window $|y| < 4$ and to have $p_t$
greater than $100\GeV$ and $80\GeV$, respectively.
For the study of the UE we used the C/A algorithm with $R=0.6$.

As a first step, we verified that the more realistic UE from MC models shows a
number of characteristic features found in our study of the toy UE.
Then, we have examined a series of observables of potential interest for
measuring at the LHC. 
 
\begin{figure}[t]
\psfig{figure=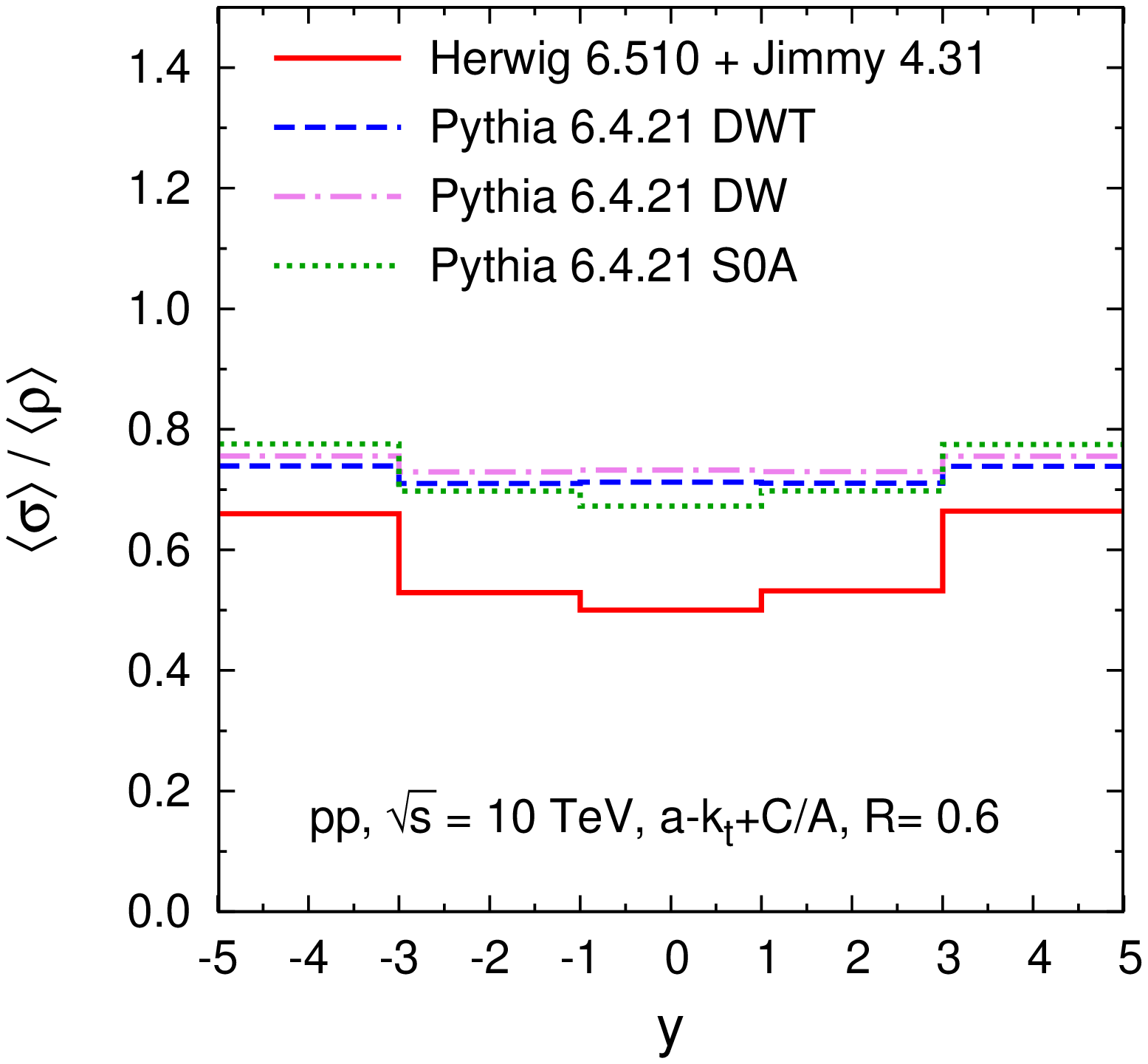,
       height=7.4cm,angle=-90}
\hfill
\psfig{figure=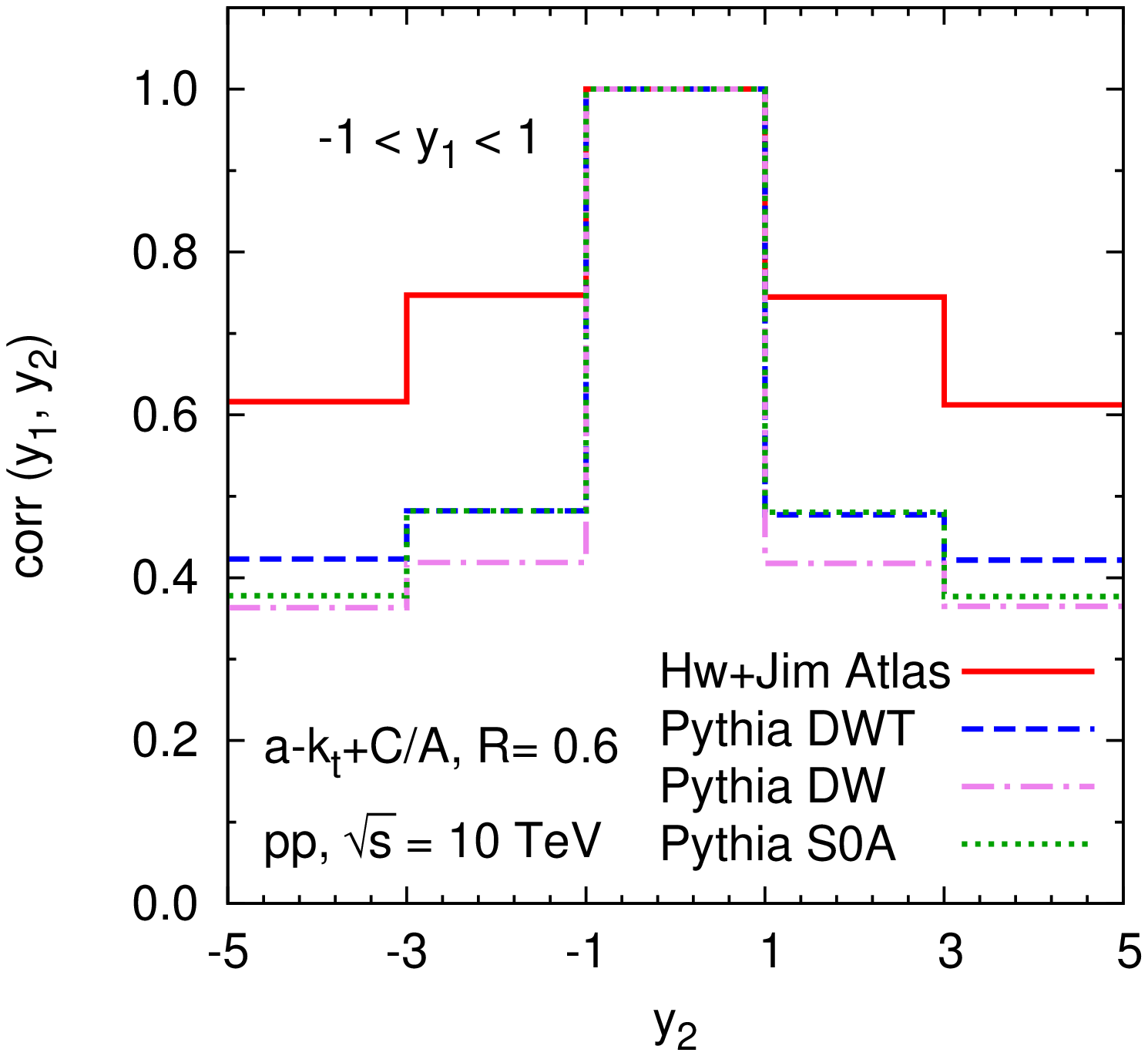,
       height=7.4cm,angle=-90}
\caption{
    Left: Average intra-event fluctuation $\lan \sigma \ran$ from series of MC
    models as function of rapidity. 
    Right: Correlations of $\rho(y_2)$ with $\rho(y_1)$, as function
    of $y_2$ for $y_1$ in the bin $-1<y<1$.
\label{fig:fluct-cor}}
\end{figure}

The main quantity, $\lan\rho\ran$, is show in
Fig.~\ref{fig:rhohist-rhomc} (right) as a function of $y$. We note that the
rapidity dependence is quite strong and the exact level of the UE
extrapolated for the LHC depends on the MC model/tune.
UE may fluctuate both from point to point within a single event and from one
event to another. The first kind of fluctuation is measured by $\lan \sigma
\ran$ and is shown in Fig.~\ref{fig:fluct-cor}~(left). We see that all the
models predict large intra-event fluctuations. Another interesting thing to note
is that in  Herwig they are nearly $40\%$ smaller than in 
Pythia. This difference is consistent with the correlation coefficients shown in
Fig.~\ref{fig:fluct-cor} (right) as a function of~$y_2$ for a single $y_1$ bin.
Altogether, the results from Figs.~\ref{fig:rhohist-rhomc}
and~\ref{fig:fluct-cor} illustrate the potential gain to be had from studying
wider variety of observables.

%%%%%%%%%%%%%%%%%%%%%%%%%%%%%%%%%%%%%%%%%%%%%%%%%%%%%%%%%%%%%%%%%%%%%%%%%%%%%%%%
\section{Conclusions}
We have carried out a twofold study devoted to the problem of measuring the
underlying event.
Using a simple toy model of UE, we have examined the methods of its
determination. Subsequently, by studying the events simulated with MC
generators, we have identified a range of important characteristics of the UE.
 
One conclusion from the toy model study is that for determinations of
averaged quantities, like~$\langle \rho \rangle$, both the traditional and the
area/median measurement methods give comparably good results.
In contrast, for event-by-event measurements and determinations of fluctuations
of the soft component, the traditional approach is affected significantly more
by the hard radiation.

Therefore, we chose the area/median method to examine more realistic UE from the
MC models. We found noticeable differences between predictions of
different generators/tunes extrapolated to LHC energy. For this reason we
advocate measuring a broader range of observables including rapidity dependence
%of $\rho$, intra-event fluctuations, $\sigma$, event-to-event fluctuations and correlations.
of $\rho$, intra- and inter-event fluctuations and correlations.

%%%%%%%%%%%%%%%%%%%%%%%%%%%%%%%%%%%%%%%%%%%%%%%%%%%%%%%%%%%%%%%%%%%%%%%%%%%%%%%%
\section*{Acknowledgments}
The original results presented here were obtained with Matteo Cacciari and Gavin
Salam. The work was supported by the French ANR under contract ANR-09-BLAN-0060
and by the Groupement d'Int\'er\^et Scientifique ``Consortium Physique des 2
Infinis'' (P2I).

%%%%%%%%%%%%%%%%%%%%%%%%%%%%%%%%%%%%%%%%%%%%%%%%%%%%%%%%%%%%%%%%%%%%%%%%%%%%%%%%

\end{document}